\documentclass{sf2a-conf2012}
\usepackage{graphicx}
\usepackage{hyperref}
\usepackage[]{natbib}
\usepackage[T1]{fontenc}
\usepackage[cyr]{aeguill}
\usepackage{epstopdf}

\def\BibTeX{{\rm B\kern-.05em{\sc i\kern-.025em b}\kern-.08em
    T\kern-.1667em\lower.7ex\hbox{E}\kern-.125emX}}
\bibpunct{(}{)}{;}{a}{}{,}  

\begin{document}

\TitreGlobal{SF2A 2012}


\title{The magnetic coupling of planets and small bodies with a pulsar's wind}

\runningtitle{planets and small bodies in a pulsar's wind}

\author{F. Mottez}\address{LUTH, Obs. de Paris, CNRS, Universit\'e Paris Diderot, 91190 Meudon}

\author{J. Heyvaerts}\address{Obs. de Strasbourg, associ\'e au LUTH}



\setcounter{page}{237}


\maketitle


\begin{abstract}
We investigate the electromagnetic interaction of a relativistic stellar wind with a planet or a smaller body in orbit around a pulsar.
This may be relevant to objects such as PSR B1257+12 and PSR B1620-26 that are expected to hold a planetary system, or to pulsars with suspected asteroids or comets. Most models of pulsar winds predict that, albeit highly relativistic, they are slower than Alfv\'en waves. In that case, a pair of stationary Alfv\'en waves, called Alfv\'en wings (AW), is expected to form on the sides of the planet. The wings expand far into the pulsar's wind and they could be strong sources of radio emissions. The Alfv\'en wings would cause a significant drift over small bodies such as asteroids and comets.
\end{abstract}



\section{Introduction}
Two pulsars with planets have been discovered, in 1992 and 1993 \citep{Wolszczan_1992,Thorsett_1993}. In particular PSR B1257+12 hosts three planets at distances of the order of the astronomical unit.
A pulsar is a neutron star with a fast spin (period $P \sim 0.001-5$ s) and surrounded by a magnetosphere that is a powerful source of radio and/or high energy emissions. (We use also the frequency $\Omega_*=2 \pi/P_*$).
Precise timing of pulsars show that they spin down at a rate that is typically 
$\dot P \sim 10^{-15}-10^{-19} P$.
The dissipation  of energy of PSR B1257+12 associated to the spin down is $\dot E_{rotation} =- M_I \Omega_* \dot \Omega_* =4 \pi^2 M_I \dot P / P^3= 2. \times 10^{27}$ W, where $M_I$ is their momentum of inertia. It can be compared to the planet's gravitational energy $E_G = {G M_* M_p}/{2 a }  = 4.  \times 10^{32}$  J. If the flux dissipated by the spin down 
was captured through the planetary radius, the gravitational energy of the  planet PSR 1257+12 "a" would be dissipated in only $ 8. \times 10^6$ years. Even with an inefficient coupling mechanism, can we expect heating, plasma acceleration and  radio emissions from the planet, or a modification of its orbit ?

\section{A planet in the wind}
The planets (with an orbital period $P_{orb} >$ a few mn) are far beyond the light cylinder; they orbit in the pulsar's wind.
This pulsar wind is an almost radial expansion of ultrarelativistic (Lorentz factor $\gamma \sim 10^2-10^7$.) and underdense plasma.
Most pulsar wind models \citep{Michel_1969,Contopoulos_1999,Spitkovsky_2006,Kirk_2009,Petri_2012a} converge on the following facts:
 This wind is highly magnetized ($B^2 >> \mu_0 \rho \gamma$),
the magnetic field is quasi-azimutal ($B \sim B_\phi >> B_{poloidal}$).
It is possibly sub-Alfv\'enic ($v_W \sim c$, $V_A \sim c$, until 100's of $r_{LC}$,  (models) $v_W < V_A$). 
This last fact means that the planet is not preceded by a shock wave, but it is directly connected to the wind. 

\section{An electromagnetic wake behind the planet: Alfv\'en wings}
Then, making simple hypothesis, it is possible to develop a theory of the interaction of the planet with the wind. The hypothesis are 
(1) the plasma is locally uniform and incompressible (OK at 1st order \citep{Wright_1990}).
 It can be non-relativistic \citep{Neubauer_1980} or relativistic flow \citep{Mottez_2011_AWW,mottez_2012c}. 
(2) Alfv\'en waves (MHD waves carrying mainly  magnetic fluctuations) propagate almost only along the magnetic field
(3) the planet is a conducting body (maybe because of a ionosphere)
	
\begin{figure}[ht!]
 \centering
 \includegraphics[width=0.38\textwidth,clip]{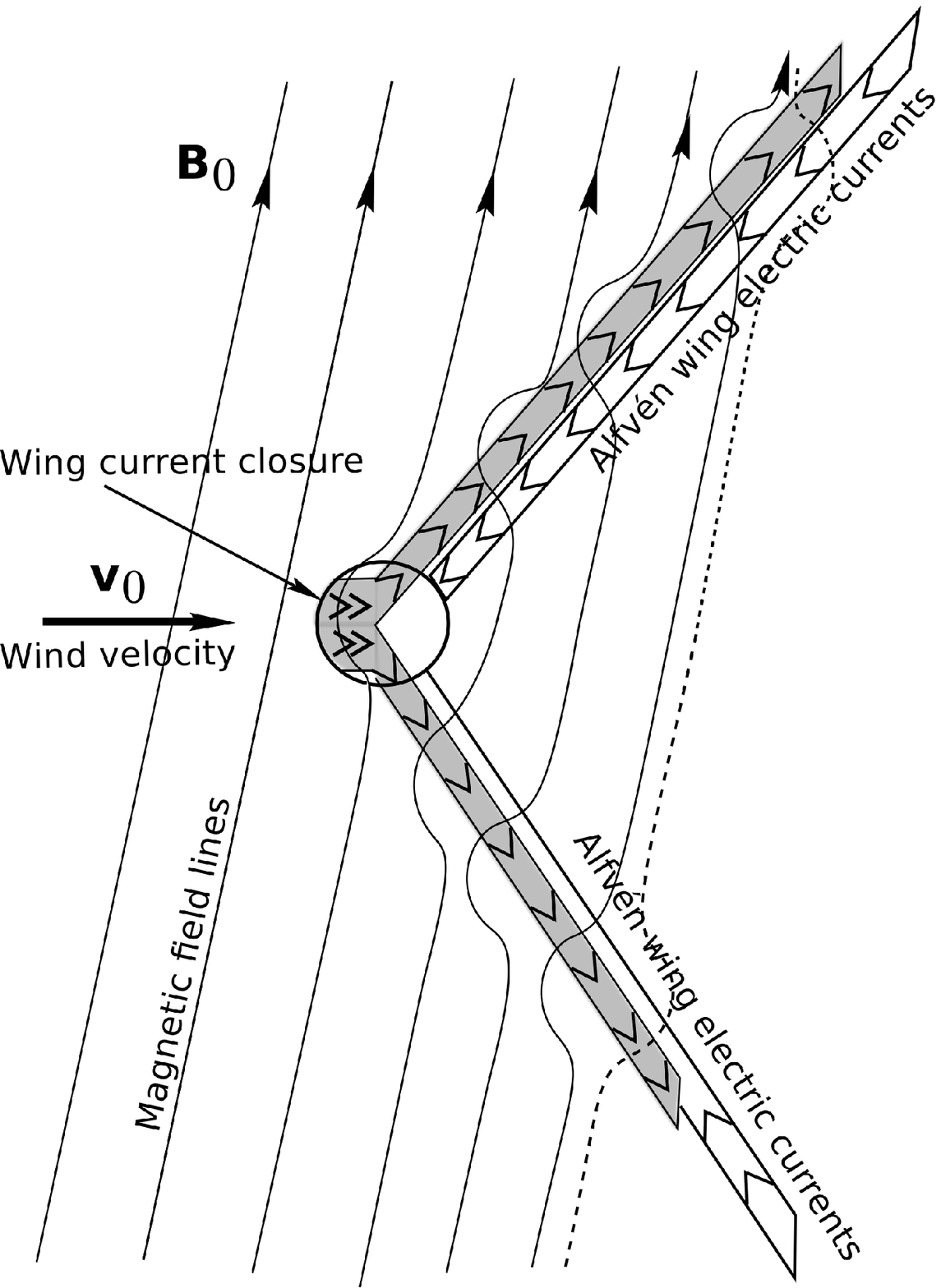}
  \includegraphics[width=0.48\textwidth,clip]{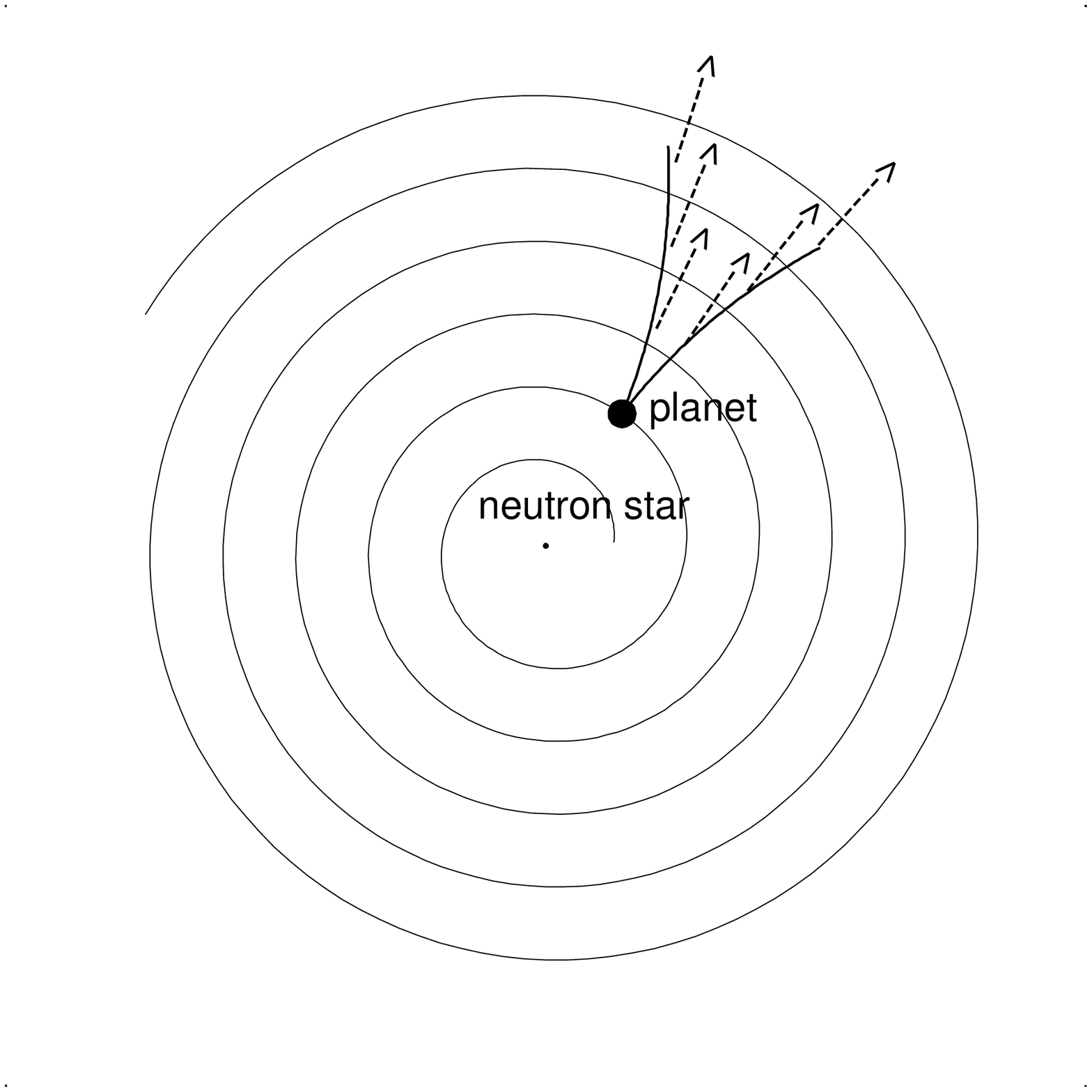}%

  \caption{\textit{Left:} The engine of the Alfv\'en wing is a unipolar inductor. The unperturbed wind's magnetic field ${\bf B}_0$ and  velocity ${\bf v}_0$ are almost, but not exactly, perpendicular. The electric field ${\bf E}_0$ created by the unipolar inductor is perpendicular to these two vectors; it induces an electric current (of density ${\bf j}$) along the body. This current then goes into the interplanetary medium, forming two structures, each of them made of an outwards and an inwards flow. The current density ${\bf j}$ flowing along the planet is the cause of a ${\bf j} \times {\bf B}$ force density. Right: general shape of the Alfv\'en wings on a larger scale. The spiral is the shape of a magnetic field line. The arrows show the directions of emissions of the radio waves if they are emitted by a source that is propagated by the wind.}
  \label{mottez:fig1}
\end{figure}

	Then,
	a potential drop is established by the plasma flow across the planet.
	The potential drop induces a current. This current is closed along the magnetic field lines, by two stationary Alfv\'en waves called Alfv\'en wings.
	This structure is a kind of electric circuit, sketched on see the left-hand side of Fig. \ref{mottez:fig1}. Its conductivity along the planet is those of the planet. Its conductivity $\Sigma_A$ along the two ribbons of current that extend into space (the Alfv\'en wings) is, in the highly relativistic case, approximately equal to the vacuum conductivity
	\begin{equation}
	\Sigma_A \sim 1/\mu_0 c \sim 0.09 \mbox{ S}.
	\end{equation}
Then, it is possible \citep{Neubauer_1980,Mottez_2011_AWW} to estimate the total current circulating in the Alfv\'en wings
\begin{equation}
I_{AW}=4 (E_0 - E_i) R_P \Sigma_A,
\end{equation}
where $E_0$ is the mean electric field associated to the unipolar inductor, $E_i$ due to the planet's resistivity (that is low), $R_P$ is the planetary radius, or more generally, the body's radius. 
For a $P=10$ms PSR and an Earth-like planet at 0.2 UA, $I_{AW} \sim 10^{9}$ A.
It is larger for a standard pulsar (of period $P=1$s), $I_{AW} \sim 10^{11}$ A. For comparison, the pulsar's current i.e. the Goldreich-Julian current that powers the whole magnetosphere is $I_{GJ}= 2. \times 10^{11}$ A. For standard pulsars, the two currents are comparable !

\section{Radio emissions ?}
Of course, these strong and localized ribbons of electric current are probably unstable. Consequently, they may constitute powerful sources of radio emissions. A priori, there are two possible regions of emission (it is nothing more than a conjecture)
(1) a radio source near the planet. Then,
	the source does not move in our frame of reference and the emissions might be similar to "usual" planetary radio emissions, generally associated to instabilities with $J_\parallel$, as it happens with the couple Jupiter-Io \citep{Queinnec_1998,mottez_2008_b,mottez_2009_b}. It would be difficult to predict the angles of emission. 
	(2) The  radio source is  along the Alfv\'en wing, and it is convected with the pulsar's wind. Then, the  source is moving fast $V \sim c$ and the emission is beamed by the relativistic aberration (see the right-hand side of Fig. \ref{mottez:fig1}). In that case, the radiation would be well collimated, and it might be observable, under lucky circumstances \citep{Mottez_2011_Graz}.

\section{Alfv\'en wings have a large influence on comets and asteroid's orbits}

\begin{figure}[ht!]
 \centering
 \includegraphics[width=0.43\textwidth,clip]{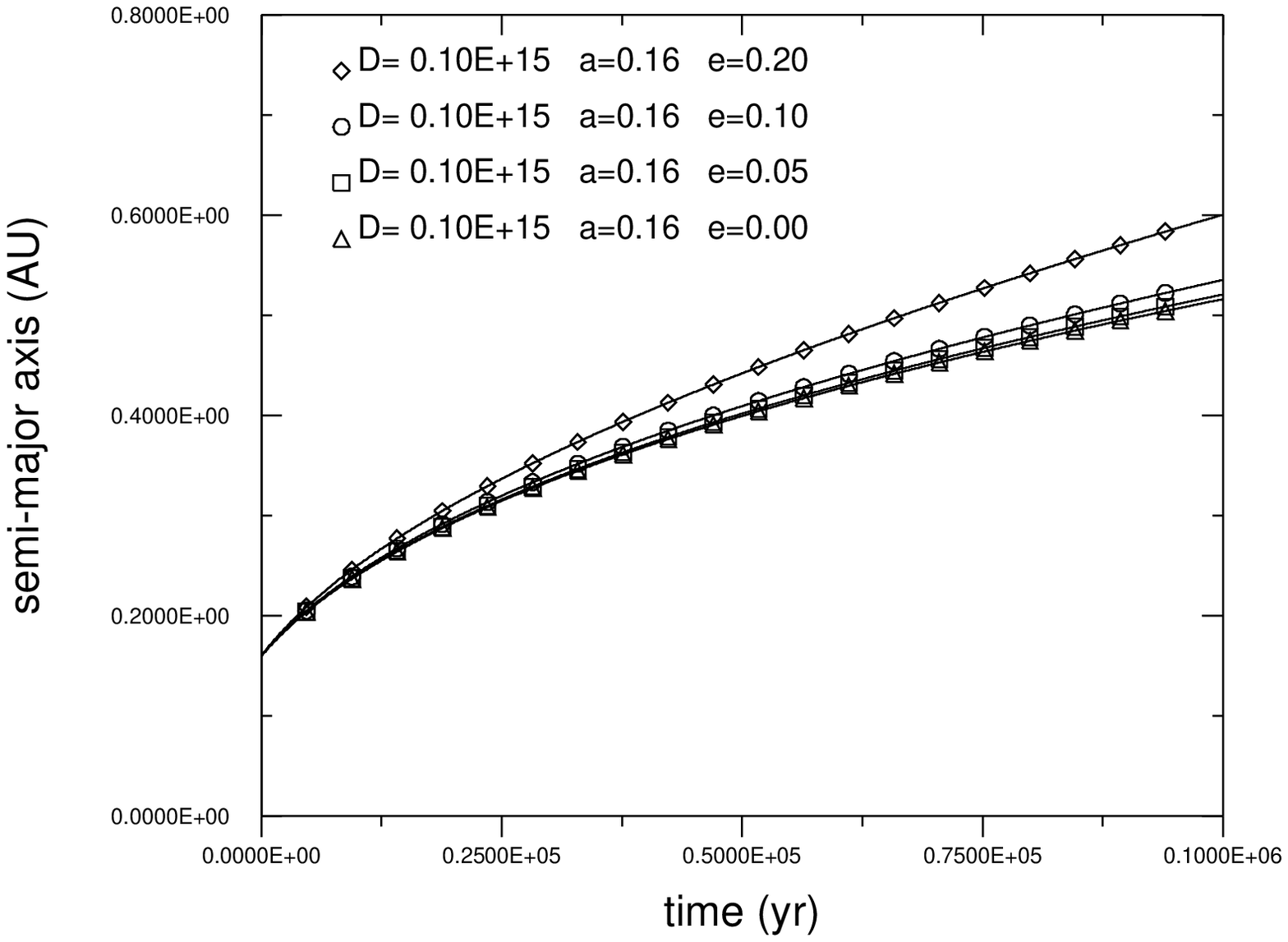}
 \includegraphics[width=0.43\textwidth,clip]{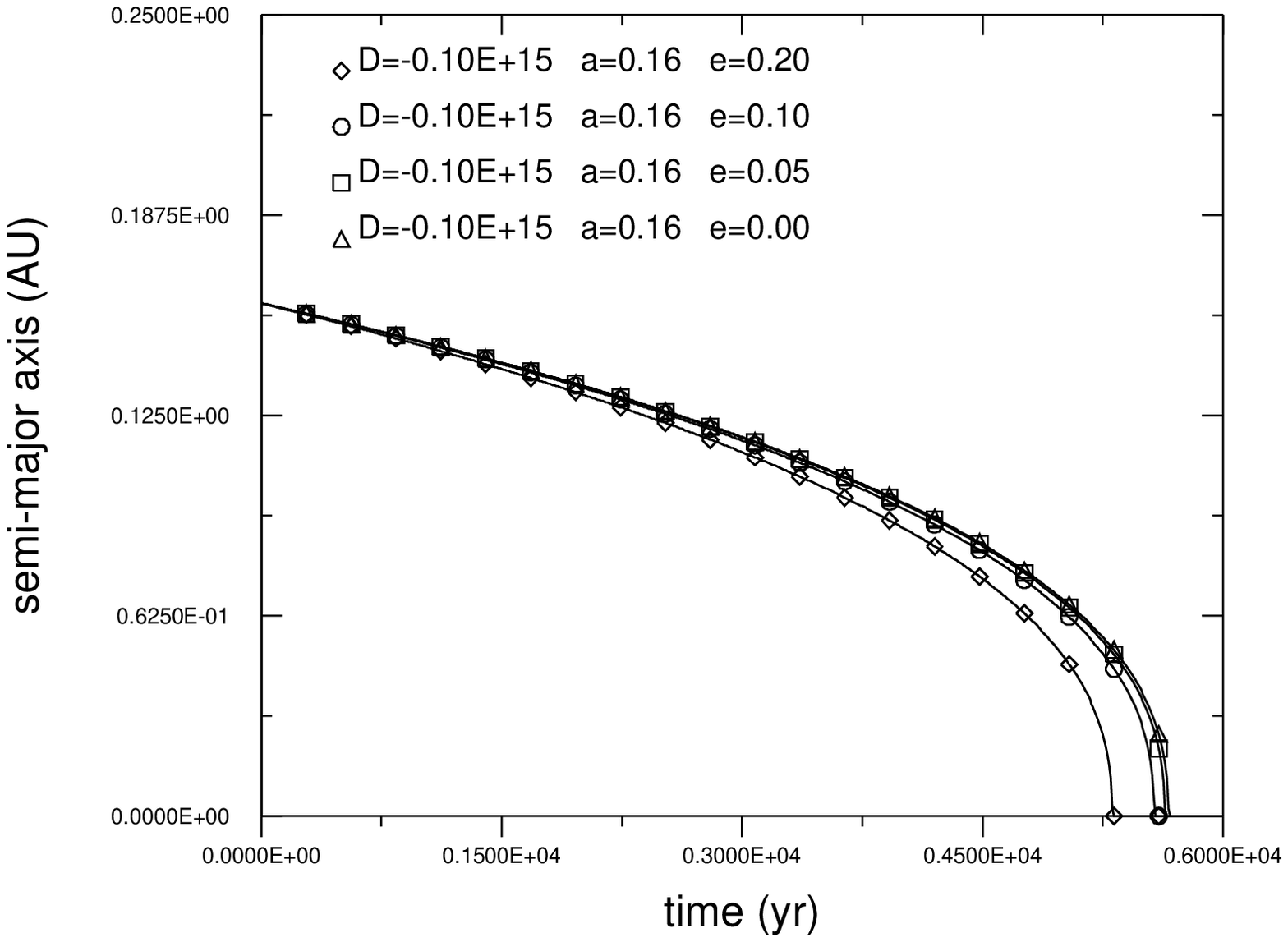}%

  \caption{ Time-evolution of the semi-major axis of a 1km sized asteroid under the influence of the magnetic thrust. The initial distance $a$ is 0.16 UA.  The different curves are traced for various values of the orbit eccentricity $e$.  \textit{Left:} The orbital spin is \textit{parallel} to the neutron star spin axis. In that case, the semi-major axis increases. The time axis covers 100,000 years. \textit{Left:} The orbital spin is \textit{anti-parallel} to the neutron star spin axis. The semi-major axis decreases. The time axis covers only 10,000 years.
  \textit{Right:} }
  \label{mottez:fig2}
\end{figure}

In the Alfv\'en wing electric circuit, the current density $\bf{j}$ flowing along the planet, crossed with the ambient magnetic field, is the cause of a $\bf{j} \times \bf{B}$ force density.
The force has a component that is tangential to the orbit.
It gives a non Keplerian contribution to the motion. This force and its effect on the orbit of a body in the pulsar wind can be estimated analytically \citep{Mottez_2011_AWO}.
Then, the evolution of the orbital elements can be computed numerically. For planets
this force has a negligible contribution to the orbit. But it becomes very important when small bodies, like comets, asteroids and planetesimal are concerned. Figure \ref{mottez:fig2} shows the time evolution of the semi-major axis of a 1 km sized rocky body in orbit around a standard pulsar (pulsar period $P=1s$). The forces acts differently if the orbital spin is parallel or anti-parallel to the neutron star' spin. When they are parallel, the semi-major axis and the eccentricity tend to increase, while they are reduced for anti-parallel spins. In this last case, we see that a body can be precipitated onto the neutron star in less than 10,000 years.

This effect of the magnetic force on the objects orbiting a pulsar, especially planetesimals, may have important consequences in scenarios of planetary formation after the transformation of the star into a neutron star, from the debris of the supernova.

\begin{table*} 
\label{table_application_planetes} 
\centering 
\begin{tabular}{l r r r r  r } 
\hline\hline 
Name & $U$ (V)& $I$ (A)& $\dot E_{J max}$ (W)& $\Delta a/year$ (m) & $\Delta e /year$ \\ 
\hline 
PSR 1257+12 a &  1.1 $\times 10^{12}$ & 3.0 $\times 10^{9}$ & 2.5$\times 10^{21}$ &  0.02 & 0 \\
PSR 1257+12 b &  3,5 $\times 10^{12}$ & 9.4 $\times 10^{9}$ & 2.5$\times 10^{22}$ &  $ 1.\times 10^{-3}$ & 3 $ \times 10^{-16}$ \\
PSR 1257+12 c &  2,6 $\times 10^{12}$ & 7.0 $\times 10^{9}$ & 1.4$\times 10^{22}$ &  $ 1.\times 10^{-3}$ & 2.4$ \times 10^{-16}$ \\
\hline 
PSR 1620-26 a &  6,0 $\times 10^{11}$ & 1.5 $\times 10^{9}$ & 7$\times 10^{20}$ &   3.$\times 10^{-6}$  & 0\\
\hline 
PSR 10ms b 100 km &   2.4$\times 10^{9}$ &  6.$\times 10^{6}$ & 1.2$\times 10^{16}$ &  $ 8.\times 10^{-3}$&   6.$\times 10^{-14}$\\
PSR 10ms b 1 km   &   2.4$\times 10^{7}$ &  6.$\times 10^{4}$ & 1.2$\times 10^{12}$ &  0.8 &  6.$\times 10^{-12}$\\
\hline 
PSR 1 s b 100 km &  2.4  $\times 10^{11}$ &  6$\times 10^{8}$ & 1.2 $\times 10^{20}$ & 8$\times 10^{4}$ & 6.4 $\times 10^{-8}$ \\
PSR 1 s b 1 km   &  2.4 $\times 10^{9}$ &  6.$\times 10^{6}$ & 1.2$\times 10^{16}$ &  8.$\times 10^{5}$  & 6.$\times 10^{-6}$  \\
\hline 
\end{tabular}
\caption{Electric potential drop, total electric current associated to the Alfv\'en wing. 
Electrical energy $\dot E_{J max}$ dissipated in the Alfv\'en wing. 
Variation per (terrestrial) year of the semi-major axis. 
Variation of the eccentricity, per year, $\Delta e /year$.}
\end{table*}

\section{Conclusions}
A planet orbiting around a pulsar would be immersed in an ultrarelativistic underdense plasma flow.\\
\vspace{0.3cm}
It would behave as a unipolar inductor, with a significant potential drop along the planet.
As for Io in Jupiter's magnetosphere, there would be two stationary Alfv\'en waves (Alfv\'en wings) attached
to the planet. \\
\vspace{0.3cm}
The AW are supported by strong electric currents, comparable to those of a pulsar.
It would be a cause of strong radio emissions, with sources all along the AW, highly colimated through 
relativistic aberration. \\
\vspace{0.3cm}
There would be a chance to detect these radio-emissions from Earth.\\
The emission would be pulses as for ordinary pulsars, but highly dependent on the planet-star-observer angle, maybe one very brief sequence once every orbit.\\ 
\vspace{0.3cm}
The Alfv\'en wing exerts a force unpon the orbiting body. This force has no influence of the motion of a planet. On the time scale of millions of years, it can however affect 
  the orbit of bodies with a diameter of 100 kilometres around
  standard pulsars with a period $P \sim $1 s and a magnetic field $B \sim  10^{8}$ T.
  Kilometer-sized bodies experience drastic orbital changes on a time scale of  $10^4$ years.
\end{document}